\documentclass[a4paper, 10 pt, conference]{ieeeconf}
\IEEEoverridecommandlockouts
\usepackage{amsfonts,amsmath,amssymb} 

\newcommand{\be}{\begin{equation}}
 \newcommand{\ee}{\end{equation}}
\newcommand{\bea}{\begin{eqnarray}}
 \newcommand{\eea}{\end{eqnarray}}
 \newcommand{\beann}{\begin{eqnarray*}}
 \newcommand{\eeann}{\end{eqnarray*}}
\newcommand{\ba}{\begin{array}}
 \newcommand{\ea}{\end{array}}

\newcommand{\bC}{{\mathbb{C}}}

\usepackage{psfrag,color}
\usepackage{enumerate,cite,latexsym,graphicx}
\newtheorem{theorem}{Theorem}

\newtheorem{corollary}{Corollary}
\newtheorem{definition}{Definition}

\def\tr{\mathop{\rm Tr}\nolimits} 

\title{\LARGE \bf Notes on Coherent Feedback Control for Linear Quantum Systems
}

\author{Ian R.~Petersen %
\thanks{This work was supported by the
Australian Research Council (ARC) and the Air Force Office of Scientific
Research (AFOSR). This material is based on research sponsored by the
Air Force Research Laboratory, under agreement numbers
FA2386-09-1-4089 and FA2386-12-1-4075.  The U.S. Government is authorized to reproduce and
distribute reprints for Governmental purposes notwithstanding any
copyright notation thereon.
The views and conclusions contained herein are those of the authors
and should not be interpreted as necessarily representing the official
policies or endorsements, either expressed or implied, of the Air
Force Research Laboratory or the U.S. Government. }%
\thanks{Ian R. Petersen is with the School of  Engineering and Information Technology, 
        University of New South Wales at the Australian Defence Force Academy, Canberra ACT 2600, Australia.
         {\tt\small i.r.petersen@gmail.com} } 
}%

\begin{document}

\maketitle
\thispagestyle{empty}
\pagestyle{empty}

\begin{abstract}
This paper considers some formulations and possible approaches to the coherent LQG and $H^\infty$ quantum control problems. Some new results for these problems are presented in the case of annihilation operator only quantum systems showing that in this case, the optimal controllers are trivial controllers. 
\end{abstract}

\section{Introduction} \label{sec:intro}
In recent years, a number of papers have considered the feedback
control of systems whose dynamics are governed by the laws of quantum 
mechanics instead of classical mechanics; see e.g., \cite{JNP1,NJP1,MaP4,GJ09}.  Also, an important class of quantum system models are linear quantum stochastic systems which  describe quantum optical devices such as optical cavities, linear quantum amplifiers, and finite bandwidth squeezers; e. g. see \cite{GZ00}. 
For such linear quantum system models an important class of quantum control problems are referred to as coherent
control problems; e.g., see \cite{JNP1,NJP1,MaP3,MaP4,MAB08,ZJ11,VP4,VP5a,HM12}. These coherent quantum control problems include the coherent LQG control problem (e.g., see \cite{NJP1}) and the coherent $H^\infty$ control problem (e.g., see \cite{JNP1}). In
coherent quantum control problems, the controller itself is required
to be a quantum system. One motivation for considering such coherent
quantum control problems is that coherent controllers have the
potential to achieve improved performance since quantum measurements
inherently involve the destruction of quantum information.

In this paper, we discuss the formulation and possible approaches to the coherent LQG and $H^\infty$ quantum control problems for a class of linear stochastic quantum system models. Also, we present some new results for these problems in the case of annihilation operator only quantum systems; e.g., see \cite{MaP4}. 

\section{Quantum Systems and Physical Realizability} \label{sec:systems}
In this section, we describe the general class of quantum systems under consideration; see also \cite{JNP1,GJ09,ZJ11}. 
We consider a
collection of $n$ independent quantum harmonic oscillators. Corresponding to this collection
of harmonic oscillators is a vector of  {\em annihilation operators}
$
a = \left[\begin{array}{cccc}a_1&a_2& \ldots & a_n
\end{array}\right]^T.
$
The adjoint of the operator $a_i$ is
  denoted by $a_i^*$
and is referred to as a \emph{creation operator}. The operators
$a_i$ and $a_i^*$ are such that the following
 \emph{commutation
relations} are satisfied:
\begin{eqnarray}
\label{CCR2} 
\lefteqn{\left[\left[\begin{array}{l}
      a\\a^\#\end{array}\right],\left[\begin{array}{l}
      a\\a^\#\end{array}\right]^\dagger\right]} \nonumber \\
&=&\left[\begin{array}{l} a\\a^\#\end{array}\right]
\left[\begin{array}{l} a\\a^\#\end{array}\right]^\dagger
- \left(\left[\begin{array}{l} a\\a^\#\end{array}\right]^\#
\left[\begin{array}{l} a\\a^\#\end{array}\right]^T\right)^T\nonumber \\
&=& \Theta
\end{eqnarray}
where $\Theta$ is a Hermitian commutation matrix of the form $\Theta = T J T^\dagger$ with $J = \left[\begin{array}{cc}I & 0\\
0 & -I\end{array}\right]$ and $T =  \Delta(T_1,T_2)$. Here $\Delta(T_1,T_2)$ denotes the matrix $\left[ \ba{cc}
 T_1 & T_2 \\ T_2^\# & T_1^\#
     \ea  \right]$. Also, $^\dagger$ denotes the adjoint transpose of a vector of operators or the complex conjugate transpose of a complex matrix. In addition, $^\#$ denotes the adjoint of a vector of operators or the complex conjugate of a complex matrix. 

The quantum harmonic oscillators 
 are assumed to be
coupled to $m$ external independent quantum fields modelled by
bosonic annihilation field operators $\mathcal{A}_1,
\mathcal{A}_2,\ldots,\mathcal{A}_m$. 
For each annihilation field operator
 $\mathcal{A}_k$, there is a
corresponding creation field operator
 $\mathcal{A}_k^*$, which is
 the operator adjoint of
$\mathcal{A}_k$. The field annihilation operators are also collected
 into a vector of
operators defined as follows:
$
 \mathcal{A}=\left[\begin{array}{cccc}
\mathcal{A}_1& \mathcal{A}_2& \ldots & \mathcal{A}_m
\end{array}\right]^T.
$

In order to describe the dynamics of a quantum linear system, we first specify the {\em
Hamiltonian operator} for the quantum system which is a Hermitian
operator on $\mathcal{H}$ of the form
\[
{\bf H} =
 \frac{1}{2}\left[\begin{array}{cc} a^\dagger &
       a^T\end{array}\right] M
\left[\begin{array}{c} a \\  a^\#\end{array}\right]
\]
where $ M$ is a Hermitian matrix of the
form
\begin{equation}
\label{tildeMN}
 M= \Delta( M_1, M_2).
\end{equation}
Also, we specify the {\em coupling operator vector} for the quantum
system to be a vector of  operators of the form
\[
L = \left[\begin{array}{cc} N_1 &  N_2 \end{array}\right]
\left[\begin{array}{c} a \\  a^\#\end{array}\right]
\]
where $ N_1 \in \mathbb{C}^{m\times n}$ and $ N_2 \in
\mathbb{C}^{m\times n}$. We can write
\[
\left[\begin{array}{c}L \\ L^\#\end{array}\right] =  N
\left[\begin{array}{c} a \\  a^\#\end{array}\right],
\]
where $  N=
 \Delta(  N_1, N_2)$. These operators then lead to the following quantum stochastic differential equations (QSDEs) which describe the dynamics of the quantum system under consideration:
\begin{eqnarray}
\label{qsde3} \left[\begin{array}{l} d a\\d
a^\#\end{array}\right] &=&  F\left[\begin{array}{l} 
a\\ a^\#\end{array}\right]dt +  G
\left[\begin{array}{l} d\mathcal{A}
\\ d\mathcal{A}^{\#} \end{array}\right];  \nonumber \\
\left[\begin{array}{l} d\mathcal{A}^{out}
\\ d\mathcal{A}^{out\#} \end{array}\right] &=&
 H \left[\begin{array}{l}  a\\
a^\#\end{array}\right]dt +  K \left[\begin{array}{l}
d\mathcal{A}
\\ d\mathcal{A}^{\#} \end{array}\right],\nonumber \\
\end{eqnarray}
where
\begin{eqnarray}
\label{FGHKform}  F =\Delta(\tilde  F_1, \tilde F_2),
 &&
 G = \Delta(\tilde G_1,\tilde  G_2), \nonumber \\
 H = \Delta(\tilde H_1,\tilde H_2),
 &&  K = \Delta(\tilde K_1,\tilde K_2),
\end{eqnarray}
and
\begin{eqnarray}
\label{generalizedFGHK1}
 F &=& -\imath \Theta   M -\frac{1}{2} \Theta  N^\dagger J  N; ~~
 G = -\Theta   N^\dagger J; \nonumber \\
 H &=&  N;~~
 K = I.
\end{eqnarray}

\begin{definition}
\label{D1}
A linear quantum system of the form (\ref{qsde3}),
(\ref{FGHKform}) is {\em physically realizable} if there exist
complex matrices $\Theta= \Theta^\dagger$, $ M = 
M^\dagger$,
$ N$, such that $\Theta$ is of the form
\begin{equation}
\label{Psiform}
\Theta = TJT^\dagger
\end{equation}
where $T=\Delta(T_1,T_2)$ is non-singular, $ M$ is of the form in (\ref{tildeMN}),
and (\ref{generalizedFGHK1}) is satisfied.
\end{definition}

\begin{theorem}
\label{T1} (See \cite{ShP5}) The linear quantum system (\ref{qsde3}), (\ref{FGHKform}) is
physically realizable if and only if there exists a complex matrix
$\Theta =\Theta^\dagger$  such that $\Theta$ is
of the form in (\ref{Psiform}), and
\begin{eqnarray}
\label{physreal1} && F\Theta + \Theta  F^\dagger + 
GJ G^\dagger = 0; 
G =  -\Theta  H^\dagger J  ; 
K = I.
\end{eqnarray}
\end{theorem}

The physical
realizability of  the linear quantum system (\ref{qsde3}), (\ref{FGHKform}) is related to the
 $(J,J)$-unitary properties of the corresponding
  transfer function
matrix
\begin{equation}
\label{TF}
\Gamma(s) = \left[\begin{array}{c|c} F &  G \\
  \hline
 H &  K \end{array}\right]=  H\left(sI-
F\right)^{-1} G+ K.
\end{equation}

\begin{definition}
\label{D2} (See \cite{KIM97,ShP5})
 The transfer function $\Gamma(s)$ 
    is said to be $(J,J)$-unitary if
  $\Gamma^\sim(s) J  \Gamma(s) = J,$
   for every $s \in  \bC$.
 \end{definition}

\begin{theorem}
\label{T2} (see \cite{ShP5})
Suppose the linear quantum system
(\ref{qsde3}), (\ref{FGHKform}) is minimal, and that $\lambda_i(
F)+\lambda_j^*( F) \neq 0$ for all eigenvalues
$\lambda_i( F)$, $\lambda_j( F)$ of the matrix $ F$.
Then this linear quantum system is
 physically realizable
if and only if the following conditions hold:
\begin{enumerate}[(i)]
\item
The system transfer function matrix $\Gamma(s)$ in (\ref{TF})  is
$(J,J)$-unitary;
\item
The matrix $ K$ is of the form
$K =  I$.
\end{enumerate}
\end{theorem}

\subsection{Annihilation Operator Quantum Systems}
\label{sec:annihilation}
An important special case of the above class of linear quantum systems occurs when the QSDEs (\ref{qsde3}) can be described purely in terms of the vector of annihilation operators $a$; e.g., see \cite{MaP3,MaP4}. In this case, we consider 
Hamiltonian operators of the form
$
{\bf H} =
 a^\dagger  Ma 
$
 and coupling operator vectors of the form 
$
L = N a 
$
where $M$ is a Hermitian matrix and $N$ is a complex matrix. In this case, we replace the commutation relations (\ref{CCR2}) by the commutation relations
\begin{eqnarray}
\label{CCR3} 
\left[a,a^\dagger\right]
&=& \Theta
\end{eqnarray}
where $\Theta$ is a positive-definite commutation matrix.
Then, the corresponding QSDEs are given by
\begin{eqnarray}
\label{qsde4}  d a &=&  F
adt +  G d\mathcal{A};  \nonumber \\
d\mathcal{A}^{out} &=&
 H adt +  K 
d\mathcal{A}
\end{eqnarray}
where
\begin{eqnarray}
\label{annihilationFGHK}
 F &=&  \Theta  \left( -\imath M +\frac{1}{2}   N^\dagger   N\right); ~~
 G = -\Theta   N^\dagger ; \nonumber \\
 H &=&  N; ~~
 K = I.
\end{eqnarray}

\begin{definition}
\label{D3}
A linear quantum system of the form (\ref{qsde4}) is {\em physically realizable} if there exist
complex matrices $\Theta > 0$, $ M = 
M^\dagger$,
$ N$, such that (\ref{annihilationFGHK}) is satisfied.
\end{definition}

\begin{theorem}
\label{T3} (See \cite{MaP3}) The annihilation operator linear quantum system (\ref{qsde4}) is
physically realizable if and only if there exists a complex matrix
$\Theta > 0$  such that 
\begin{eqnarray}
\label{physreal3} && F\Theta + \Theta  F^\dagger + 
G G^\dagger = 0;
G =  -\Theta  H^\dagger; 
K = I.
\end{eqnarray}
\end{theorem}

The physical
realizability of  the annihilation operator linear quantum system (\ref{qsde4}) is related to the
lossless bounded real property of the corresponding
  transfer function
matrix
\begin{equation}
\label{TF1}
\Gamma(s) = \left[\begin{array}{c|c} F &  G \\
  \hline
 H &  K \end{array}\right]=  H\left(sI-
F\right)^{-1} G+ K.
\end{equation}

\begin{definition}
\label{D4}
 The transfer function $\Gamma(s)$ 
    is said to be lossless bounded real  if all of the poles of $\Gamma(s)$ are in the open left half of the complex plane and 
  $\Gamma^\sim(s)  \Gamma(s) = I$
   for every $s \in  \bC$.
 \end{definition}

\begin{theorem}
\label{T4} (See \cite{MaP3})
A minimal annihilation operator linear quantum system (\ref{qsde4}) is
physically realizable if and only if 
the system transfer function matrix $\Gamma(s)$ in (\ref{TF1}) is lossless bounded real and the  matrix $ K$ is of the form
$K =  I$.
\end{theorem}

\section{Coherent Feedback Control}
\label{sec:coherent}
We now consider problems of coherent feedback control for linear quantum systems. In these problems, both the plant and the controller are linear quantum systems. Moreover, in the plant, the input fields to the system are divided into two types, control input fields and vacuum noise input fields. Hence, we re-write the system (\ref{qsde3}) in the following form.

\noindent
{\bf Plant:}
\begin{eqnarray}
\label{plant} 
\left[\begin{array}{l} d a\\d
a^\#\end{array}\right]
&=&  F\left[\begin{array}{l} 
a\\ a^\#\end{array}\right]dt 
+\left[\begin{array}{ll} G_{w1} &  G_{w2}\end{array}\right]
\left[\begin{array}{l}  d\mathcal{W} \\ d\mathcal{W}^\# 
\end{array}\right]
\nonumber \\&&
+ \left[\begin{array}{ll} G_{u1} & G_{u2}\end{array}\right] 
\left[\begin{array}{l}d\mathcal{U} \\ d\mathcal{U}^\#\end{array}\right]; \nonumber \\
\left[\begin{array}{l} d\mathcal{Y} \\d\mathcal{Y}^{\#}\end{array}\right]
 &=&
\left[\begin{array}{l} H_1 \\  H_2 \end{array}\right] \left[\begin{array}{l}  a\\
a^\#\end{array}\right]dt \nonumber \\
 &&+ \left[\begin{array}{ll} K_{11} &  K_{12} \\
K_{21} & K_{22}\end{array}\right]
\left[\begin{array}{l}  d\mathcal{W} \\ d\mathcal{W}^\# 
\end{array}\right]
\end{eqnarray}
where $\mathcal{U}$ denotes  the control input field,  $\mathcal{W}$ denotes  the noise input field, and $\mathcal{Y}$ denotes  the output field. In coherent feedback control, it is assumed that the output field is an input field for another quantum linear system, which is the {\em controller} and the control input field is an output field for the controller system. Note that it is assumed that there is no direct feedthrough from the control input field $\mathcal{U}$ to the output field $\mathcal{Y}$. The transfer function matrix corresponding to the plant (\ref{plant}) will be denoted
$\Gamma_{P}(s) = \left[\begin{array}{ll}\Gamma_{P1}(s)& \Gamma_{P2}(s)\end{array}\right]$.

It will be assumed that the plant is physically realizable. That is, it is assumed that the plant (\ref{plant}) can be augmented with an additional (unused) output to obtain a quantum linear system of the form (\ref{qsde3}) which is physically realizable. This augmented plant is defined as follows:
\begin{eqnarray}
\label{augplant} 
\left[\begin{array}{l} d a\\d
a^\#\end{array}\right]
 &=&  F\left[\begin{array}{l} 
a\\ a^\#\end{array}\right]dt \nonumber \\
\lefteqn{+  \left[\begin{array}{llll}G_{w1}& G_{u1}& G_{w2}& G_{u2} \end{array}\right]
\left[\begin{array}{l} d\mathcal{W}\\d\mathcal{U}
\\ d\mathcal{W}^{\#}\\ d\mathcal{U}^{\#} \end{array}\right];}
\nonumber \\
\left[\begin{array}{l} d\mathcal{Y}
\\ d\mathcal{\tilde Y} \\
d\mathcal{Y}^{\#} \\
d\mathcal{\tilde Y}^{\#}
\end{array}\right]
 &=&
 \left[\begin{array}{c}H_1\\\tilde H_1\\H_2\\\tilde H_2\end{array}\right] \left[\begin{array}{l}  a\\
a^\#\end{array}\right]dt \hspace{2.5cm}\nonumber \\
\lefteqn{ + \left[\begin{array}{llll} K_{11} & 0 & K_{12} & 0\\
\tilde K_{11} & \bar K_{11} & \tilde K_{12} & \bar K_{12}\\
 K_{21} & 0 & K_{22} & 0\\
\tilde K_{21} & \bar K_{21} & \tilde K_{22} & \bar K_{22}\\
\end{array}\right] \left[\begin{array}{l} d\mathcal{W}\\d\mathcal{U}
\\ d\mathcal{W}^{\#}\\ d\mathcal{U}^{\#} \end{array}\right].}\nonumber \\
\end{eqnarray}
 Since physical realizability requires that the direct feedthrough matrix for this system is the identity, we  have $K_{21}= 0$, $\tilde K_{21}= 0$, $\bar K_{21}= 0$, $K_{12}= 0$, and $\bar K_{12}= 0$.

The  system describing the controller is defined as follows:

\noindent
{\bf Controller:}
\begin{eqnarray}
\label{controller} 
\left[\begin{array}{l} d a_c\\d
a_c^\#\end{array}\right]
&=&  F_c\left[\begin{array}{l} 
a_c\\ a_c^\#\end{array}\right]dt 
\nonumber \\&&
+\left[\begin{array}{ll}G_{cw1} & G_{cw2}\end{array}\right]
\left[\begin{array}{l} 
d\mathcal{\tilde W} \\ d\mathcal{\tilde W}^\# 
\end{array}\right]  \nonumber \\
&&+ \left[\begin{array}{ll} G_{cy1}& G_{cy2}\end{array}\right]
\left[\begin{array}{l} 
 d\mathcal{Y} \\  d\mathcal{Y}^\#
\end{array}\right]; \nonumber \\
\left[\begin{array}{l} 
 d\mathcal{U} \\ d\mathcal{U}^{\#}
\end{array}\right]
 &=&
\left[\begin{array}{l}  H_{c1} \\ H_{c2}\end{array}\right]
\left[\begin{array}{l}  a_c\\
a_c^\#\end{array}\right]dt \nonumber \\
&& + \left[\begin{array}{ll} K_{cw11} &   K_{cw12} \\
K_{cw21} & K_{cw22}
\end{array}\right] 
\left[\begin{array}{l} 
d\mathcal{\tilde W} \\ d\mathcal{\tilde W}^\# 
\end{array}\right] \nonumber \\
&& + \left[\begin{array}{ll} K_{cy11} &   K_{cy12} \\
K_{cy21} & K_{cy22}
\end{array}\right] 
\left[\begin{array}{l} 
d\mathcal{Y} \\ d\mathcal{Y}^\# 
\end{array}\right] 
\end{eqnarray}
where $a_c$ is the controller system annihilation operator, $\mathcal{Y}$ denotes the  input field to the controller, which is the plant output field,  $\mathcal{\tilde W}$ denotes  the controller noise input field, and $\mathcal{U}$ denotes  the output field of the controller, which is the input field for the plant. The corresponding controller transfer function matrix is denoted $\Gamma_c(s)=\left[\begin{array}{ll}\Gamma_{c1}(s)& \Gamma_{c2}(s)\end{array}\right]$.

The controller is also assumed to be physically realizable. That is, it is assumed that the controller (\ref{controller}) can be augmented with an additional (unused) output to obtain a quantum linear system of the form (\ref{qsde3}) which is physically realizable. This augmented controller is defined as follows:
\begin{eqnarray*}
\left[\begin{array}{l} 
d a_c\\ da_c^\#
\end{array}\right]
 &=&  F_c\left[\begin{array}{l} 
a_c\\ a_c^\#\end{array}\right]dt \nonumber \\
\lefteqn{+  \left[\begin{array}{llll}G_{cw1}& G_{cy1}& G_{cw2}& G_{cy2} \end{array}\right]
\left[\begin{array}{l} d\mathcal{\tilde W}\\d\mathcal{Y}
\\ d\mathcal{\tilde W}^{\#}\\ d\mathcal{Y}^{\#} \end{array}\right];}
\nonumber \\
\left[\begin{array}{l} d\mathcal{U}
\\ d\mathcal{\tilde U} \\
d\mathcal{U}^{\#} \\
d\mathcal{\tilde U}^{\#}
\end{array}\right]
 &=&
 \left[\begin{array}{c}H_{c1}\\\tilde H_{c1}\\H_{c2}\\\tilde H_{c2}\end{array}\right] \left[\begin{array}{l}  a_c\\
a_c^\#\end{array}\right]dt \hspace{3.5cm}
\end{eqnarray*}
\begin{eqnarray}
\label{augcontroller} 
 + \left[\begin{array}{llll} K_{cw11} & K_{cy11} & K_{cw12} & K_{cy12}\\
\tilde K_{cw11} & \bar K_{cy11} & \tilde K_{cw12} & \bar K_{cy12}\\
 K_{cw21} &  K_{cy21} & K_{cw22} &  K_{cy21}\\
\tilde K_{cw21} & \bar K_{cy21} & \tilde K_{cw22} & \bar K_{cy22}\\
\end{array}\right] \left[\begin{array}{l} d\mathcal{\tilde W}\\d\mathcal{Y}
\\ d\mathcal{\tilde W}^{\#}\\ d\mathcal{Y}^{\#} \end{array}\right].
\end{eqnarray}
 Since physical realizability requires that the direct feedthrough matrix for this system is the identity, we must have $K_{cw21}= 0$, $\tilde K_{cw21}= 0$, $\bar K_{cy21}= 0$, $K_{cw12}= 0$, $K_{cy12}= 0$ and $\bar K_{cy12}= 0$. The corresponding augmented controller transfer function matrix  is denoted $\tilde \Gamma_c(s)$.

When the coherent controller is connected to the quantum linear system plant, we obtain the  closed loop linear quantum system. This  system is described by the following QSDEs.

\noindent
{\bf Closed Loop System:}
\begin{eqnarray}
\label{closedloop} 
\left[\begin{array}{l} d a\\d
a^\#\\d a_c\\ da_c^\#\end{array}\right]
 &=& \left[\begin{array}{ll}  F+G_uK_{cy}H & G_u H_c \\G_{cy} H & F_c 
\end{array}\right]
\left[\begin{array}{l} 
a\\ a^\# \\a_c\\ a_c^\#
\end{array}\right]dt \nonumber \\
\lefteqn{+  \left[\begin{array}{ll}G_{w}+G_uK_{cy}K& G_{u}K_{cw}\\ G_{cy}K & G_{cw} \end{array}\right]
\left[\begin{array}{l} d\mathcal{W}\\ d\mathcal{W}^{\#}\\
d\mathcal{\tilde W}\\ d\mathcal{\tilde W}^{\#} \end{array}\right]}\nonumber \\
\end{eqnarray}
where
\begin{eqnarray*}
G_u &=& \left[\begin{array}{ll}G_{u1} & G_{u2} \end{array}\right]; ~~
H_c = \left[\begin{array}{l}H_{c1} \\  H_{c2} \end{array}\right]; \nonumber \\
G_{cy} &=& \left[\begin{array}{ll}G_{cy1} & G_{cy2} \end{array}\right]; ~~
H = \left[\begin{array}{l}H_{1} \\  H_{2} \end{array}\right]; \nonumber \\
G_w &=& \left[\begin{array}{ll}G_{w1} & G_{w2} \end{array}\right]; ~~
K_{cw} = \left[\begin{array}{ll}K_{cw11} & K_{cw12} \\ K_{cw21} & K_{cw22} \end{array}\right]; \nonumber \\
G_{cw} &=& \left[\begin{array}{ll}G_{cw1} & G_{cw2} \end{array}\right]; ~~
K = \left[\begin{array}{ll}K_{11} & K_{12} \\ K_{21} & K_{22} \end{array}\right]; \nonumber \\
K_{cy} &=& \left[\begin{array}{ll}K_{cy11} & K_{cy12} \\ K_{cy21} & K_{cy22} \end{array}\right].
\end{eqnarray*}

\noindent
{\bf Cost Output:}
The coherent LQG control problem and the coherent $H^\infty$ control problem are both defined in terms of the following  cost output for the plant (\ref{plant}):
\begin{eqnarray}
\label{costoutput} 
d\mathcal{Z}
 &=&
 C \left[\begin{array}{l}  a\\
a^\#\end{array}\right]dt  +  D \left[\begin{array}{l} d\mathcal{U} \\ d\mathcal{U}^{\#}\end{array}\right].
\end{eqnarray}
Note that  no physical realizability restrictions are placed on this output equation at this stage. The transfer function matrix from the two inputs of the plant (\ref{plant}) to the performance output $\mathcal{Z}(t)$ will be denoted
$\Gamma_{z}(s) = \left[\begin{array}{ll}\Gamma_{z1}(s)& \Gamma_{z2}(s)\end{array}\right]$. The resulting closed-loop transfer function matrix from the noise inputs to the cost output is then calculated to be
\begin{eqnarray*}
\lefteqn{\Gamma_{cl}(s) =}\nonumber \\
&& \left[\begin{array}{ll} 
\Gamma_{z1}(s)+M(s)\Gamma_{c2}(s)\Gamma_{P1}(s) &
M(s)\Gamma_{c1}(s)
\end{array}\right]
\end{eqnarray*}
where $M(s) = \Gamma_{z2}(s)\left(I-\Gamma_{c2}(s)\Gamma_{P2}(s)\right)^{-1}$.

\noindent
{\bf Coherent Quantum LQG Control:}
Using Theorem \ref{T2}, a coherent quantum LQG control problem can be formulated purely in terms of transfer function matrices as follows:
\[
\min_{\Gamma_c(s)} \|\Gamma_{cl}(s)\|_2
\]
subject to the constraints that the closed-loop system is internally stable, $\tilde \Gamma_c(s)$ is $(J,J)$-unitary, and $p_i+p_j^* \neq 0$ for all poles
$p_i$, $p_j$ of the transfer function matrix $\Gamma_c(s)$. Here $\|\Gamma_{cl}(s)\|_2$ denotes the $H_2$ norm of the transfer function matrix $\Gamma_{cl}(s)$:
\[\|\Gamma_{cl}(s)\|_2^2 := \frac{1}{2 \pi} \int_{-\infty}^{\infty}\tr[\Gamma_{cl}(i\omega)\Gamma_{cl}^\dagger(i\omega)]d\omega;
\]
e.g., see Section 3.3.3 of \cite{GL95}. Here, we assume that the  transfer function matrices $\Gamma_{Z2}(s)\Gamma_{c1}(s)$ and $\Gamma_{Z2}(s)\Gamma_{c2}(s)\Gamma_{P1}(s)$ are strictly proper. Also, it follows from the form of (\ref{costoutput}) that the transfer function matrix $\Gamma_{z1}(s)$ is strictly proper. These conditions ensure that the quantity $\|\Gamma_{cl}(s)\|_2^2$ is finite provided that the closed loop system is stable. 

\noindent
{\bf Coherent Quantum $H^\infty$ Control:}
Similarly to the above, we can also formulate a coherent $H^\infty$ control problem purely in terms of transfer function matrices as follows:
\[
\min_{\Gamma_c(s)} \|\Gamma_{cl}(s)\|_\infty
\]
subject to the constraints that the closed-loop system is internally stable, $\tilde \Gamma_c(s)$ is $(J,J)$-unitary, and $p_i+p_j^* \neq 0$ for all poles
$p_i$, $p_j$ of the transfer function matrix $\Gamma_c(s)$. Note that this coherent $H^\infty$ control problem is different to the coherent $H^\infty$ control problem considered in previous papers such as \cite{JNP1,MaP4} in that we consider all quantum noise inputs as disturbances,  including the controller quantum noise inputs. 

These frequency domain formulations of the coherent quantum LQG control problem and the coherent $H^\infty$ control problem motivate numerical methods to solve these problems using the frequency domain optimization tools developed in \cite{HM97a}. These approaches are beyond the scope of the current paper but may be pursued in future research. 

\noindent
{\bf Time Domain Formulations of Coherent Quantum $H^\infty$ and LQG Control:}
To develop time domain formulations of the coherent quantum $H^\infty$ and LQG control problems, we first consider a time domain version of the physical realizability condition that $\tilde \Gamma_c(s)$ is $(J,J)$ unitary. Note that that we can write
\begin{eqnarray*}
\left[\begin{array}{l} \mathcal{U}(s)
\\ \mathcal{\tilde U}(s) \\
\mathcal{U}^{\#}(s) \\
\mathcal{\tilde U}^{\#}(s)
\end{array}\right]
 &=&\tilde \Gamma_c(s)
\left[\begin{array}{l} \mathcal{\tilde W}(s)\\\mathcal{Y}(s)
\\ \mathcal{\tilde W}^{\#}(s)\\ \mathcal{Y}^{\#}(s) \end{array}\right]
\end{eqnarray*}
for the following classical LTI system corresponding to the quantum system (\ref{augcontroller}) defined as follows:
\begin{eqnarray*}
\left[\begin{array}{l} 
\dot a_c\\ \dot a_c^\#
\end{array}\right]
 &=&  F_c\left[\begin{array}{l} 
a_c\\ a_c^\#\end{array}\right] \hspace{3cm}\nonumber \\
\lefteqn{+  \left[\begin{array}{llll}G_{cw1}& G_{cy1}& G_{cw2}& G_{cy2} \end{array}\right]
\left[\begin{array}{l} \mathcal{\tilde W}\\\mathcal{Y}
\\ \mathcal{\tilde W}^{\#}\\ \mathcal{Y}^{\#} \end{array}\right];}
\nonumber \\
\left[\begin{array}{l} \mathcal{U}
\\ \mathcal{\tilde U} \\
\mathcal{U}^{\#} \\
\mathcal{\tilde U}^{\#}
\end{array}\right]
 &=&
 \left[\begin{array}{c}H_{c1}\\\tilde H_{c1}\\H_{c2}\\\tilde H_{c2}\end{array}\right] \left[\begin{array}{l}  a_c\\
a_c^\#\end{array}\right]\hspace{3cm}
\end{eqnarray*}
\begin{eqnarray}
\label{augcontrollerclassical} 
+ \left[\begin{array}{llll} K_{cw11} & K_{cy11} & K_{cw12} & K_{cy12}\\
\tilde K_{cw11} & \bar K_{cy11} & \tilde K_{cw12} & \bar K_{cy12}\\
 K_{cw21} &  K_{cy21} & K_{cw22} &  K_{cy21}\\
\tilde K_{cw21} & \bar K_{cy21} & \tilde K_{cw22} & \bar K_{cy22}\\
\end{array}\right] \left[\begin{array}{l} \mathcal{\tilde W}\\\mathcal{Y}
\\ \mathcal{\tilde W}^{\#}\\ \mathcal{Y}^{\#} \end{array}\right].
\end{eqnarray}
In this classical system, all quantities are complex vectors or matrices. 
Then as in \cite{KIM97}, we can write
\begin{eqnarray*}
\left[\begin{array}{l} \mathcal{U}(s)
\\ \mathcal{\tilde U}(s) \\ \mathcal{\tilde W}^{\#}(s)\\ \mathcal{Y}^{\#}(s)
\end{array}\right]
 &=& \mbox{CHAIN}^{-1}\left(\tilde\Gamma_c(s)\right)
\left[\begin{array}{l}  \mathcal{U}^{\#}(s) \\
\mathcal{\tilde U}^{\#}(s)\\
\mathcal{\tilde W}(s)\\\mathcal{Y}(s) \end{array}\right]
\end{eqnarray*}
where $\mbox{CHAIN}^{-1}\left(\tilde\Gamma_c(s)\right)$ denotes the inverse of the chain scattering representation of the transfer function matrix $\tilde\Gamma_c(s)$. 
Then the transfer function matrix $\tilde\Gamma_c(s)$ is  $(J,J)$ unitary if and only if the transfer function matrix $\mbox{CHAIN}^{-1}\left(\tilde\Gamma_c(s)\right)$ is lossless bounded real; e.g., see Lemma 4.4 of \cite{KIM97} and the proof of Theorem 2 in \cite{ShP5}. Furthermore, it follows from Lemma 2.6.1 of \cite{AV73} that  the transfer function matrix $\mbox{CHAIN}^{-1}\left(\tilde\Gamma_c(s)\right)$ is lossless bounded real if and only if 
\begin{eqnarray}
\label{timedomainPR1}
\int_0^T\left(\begin{array}{l}
\|\mathcal{U}\|^2 +\|\mathcal{\tilde U}\|^2 +\|\mathcal{\tilde W}^{\#}\|^2+\|\mathcal{Y}^{\#}\|^2\\
-\|\mathcal{U}^{\#}\|^2-\|\mathcal{\tilde U}^{\#}\|^2
- \|\mathcal{\tilde W}\|^2
-\|\mathcal{Y}\|^2
\end{array}\right)dt
\geq 0
\end{eqnarray}
for all $T > 0$ and all solutions to (\ref{augcontrollerclassical}) with zero initial condition, and 
\begin{eqnarray}
\label{timedomainPR2}
\int_0^\infty\left(\begin{array}{l}
\|\mathcal{U}\|^2 +\|\mathcal{\tilde U}\|^2 +\|\mathcal{\tilde W}^{\#}\|^2+\|\mathcal{Y}^{\#}\|^2\\
-\|\mathcal{U}^{\#}\|^2-\|\mathcal{\tilde U}^{\#}\|^2
- \|\mathcal{\tilde W}\|^2
-\|\mathcal{Y}\|^2\end{array}\right)dt
= 0
\end{eqnarray}
for all solutions to (\ref{augcontrollerclassical}) with zero initial condition satisfying
\[
\int_0^\infty\left(\begin{array}{l}
\|\mathcal{U}^{\#}\|^2+\|\mathcal{\tilde U}^{\#}\|^2
+ \|\mathcal{\tilde W}\|^2
+\|\mathcal{Y}\|^2\end{array}\right)dt < \infty.
\]

To consider the time domain version of the coherent quantum $H^\infty$ and LQG control problems, we now consider the following classical LTI system corresponding to the quantum plant (\ref{plant}) with cost output (\ref{costoutput}):
\begin{eqnarray}
\label{plantclassical} 
\left[\begin{array}{l} \dot a\\ \dot
a^\#\end{array}\right]
&=&  F\left[\begin{array}{l} 
a\\ a^\#\end{array}\right] \nonumber \\
&&+\left[\begin{array}{ll} G_{w1} &  G_{w2}\end{array}\right]
\left[\begin{array}{l}  \mathcal{W} \\ \mathcal{W}^\# 
\end{array}\right]
  \nonumber \\
&&+ \left[\begin{array}{ll} G_{u1} & G_{u2}\end{array}\right] 
\left[\begin{array}{l}\mathcal{U} \\ \mathcal{U}^\#\end{array}\right]; \nonumber \\
\left[\begin{array}{l} \mathcal{Y} \\\mathcal{Y}^{\#}\end{array}\right]
 &=&
\left[\begin{array}{l} H_1 \\  H_2 \end{array}\right] \left[\begin{array}{l}  a\\
a^\#\end{array}\right] \nonumber \\
&& + \left[\begin{array}{ll} K_{11} &  K_{12} \\
K_{21} & K_{22}\end{array}\right]
\left[\begin{array}{l}  \mathcal{W} \\ \mathcal{W}^\# 
\end{array}\right]; \nonumber \\
\mathcal{Z}
 &=&
 C \left[\begin{array}{l}  a\\
a^\#\end{array}\right]  +  D \left[\begin{array}{l} \mathcal{U} \\ \mathcal{U}^{\#}\end{array}\right].
\end{eqnarray}
In this classical system, all quantities are complex vectors or matrices. 
Then we have
\[
\mathcal{Z}(s) = \Gamma_{cl}(s)\left[\begin{array}{l}\mathcal{W}(s) \\ \mathcal{W}^\#(s) \\
\mathcal{\tilde W}(s)\\ \mathcal{\tilde W}^{\#}(s)\end{array}\right].
\]

\noindent
{\bf Time Domain Coherent Quantum LQG Control:}
We can calculate the LQG control cost function $\|\Gamma_{cl}(s)\|_2$ in the time domain as
\begin{equation}
\label{timedomainlqgcost}
\|\Gamma_{cl}(s)\|_2^2 = \sum_{k=1}^{n+\tilde n}\int_0^\infty\|\mathcal{Z}_k(t)\|^2dt
\end{equation}
where $\mathcal{Z}_k(t)$ denotes the response of the closed loop system with zero initial condition, such that the $k$th noise input is a unit impulse, and all other noise inputs are zero; e.g., see Section 6.1 of \cite{DP00}. Thus, the time domain formulation of the coherent quantum LQG control problem involves minimizing the cost in (\ref{timedomainlqgcost}) for the system (\ref{plantclassical}) over the controllers (\ref{augcontrollerclassical}), subject to the constraints defined by (\ref{timedomainPR1}), (\ref{timedomainPR2}).

\noindent
{\bf Time Domain Coherent Quantum $H^\infty$ Control:}
In the time domain,  $H^\infty$ control problems are  conveniently considered as sub-optimal problems via a game theory approach; e.g., see \cite{BB95,GL95}. That is, the condition $\|\Gamma_{cl}(s)\|_\infty < \gamma$ is equivalent to the condition 
\begin{eqnarray*}
\sup_{\mathcal{W}(\cdot),\mathcal{W}^\#(\cdot), \mathcal{\tilde W}(\cdot), \mathcal{\tilde W}(\cdot)^\#}
J_\gamma  < \infty 
\end{eqnarray*}
for all $T >  0$. Here
\[
J_\gamma = \int_0^T\left(\begin{array}{l}\|\mathcal{Z}\|^2 
-\gamma^2\left(\begin{array}{l}\|\mathcal{W}\|^2+\|\mathcal{W}^\#\|^2\\
+\| \mathcal{\tilde W}\|^2
+\| \mathcal{\tilde W}^\#\|^2
\end{array}\right)\end{array}\right)dt.
\]
Hence, we can consider the following game problem for the system (\ref{plantclassical}):
\begin{equation}
\label{Hinfgame}
\inf_{\mathcal{U}(\cdot),\mathcal{U}^\#(\cdot)} \sup_{\mathcal{W}(\cdot),\mathcal{W}^\#(\cdot), \mathcal{\tilde W}(\cdot), \mathcal{\tilde W}(\cdot)^\#}
J_\gamma  < \infty 
\end{equation}
subject to the constraints defined by (\ref{timedomainPR1}), (\ref{timedomainPR2}).

\noindent
{\bf Restriction to a Strictly Proper Controller:}
In the standard LQG and $H^\infty$ control problems, the controller is usually restricted to be strictly proper. However, in the quantum case, a strictly proper controller may not be physically realizable. However, once we have decided on the dimension of the controller noise vector, the direct feedthrough matrix in the system (\ref{augcontroller}) is restricted to be the identity matrix. This determines the direct feedthrough matrices $K_{cw}$ and $K_{cy}$ in the controller (\ref{controller}). Since these matrices are fixed, these terms in the controller (\ref{augcontroller}) can be incorporated into the plant (\ref{plant}). In addition, we assume that the dimension of the controller noise $\mathcal{\tilde W}$ is greater than or equal to the the dimension of the plant input $\mathcal{U}$ and we write 
\[
\mathcal{\tilde W} = \left[\begin{array}{l} \mathcal{\tilde W}_a \\\mathcal{\tilde W}_b\end{array}\right]
\]
where the dimension of  $\mathcal{\tilde W}_a$ is equal to the dimension of $\mathcal{U}$. This leads to the following modified plant and controller classical systems:

\noindent
{\bf Modified Plant:}
\begin{eqnarray}
\label{modifiedplant} 
\left[\begin{array}{l} \dot a\\ \dot
a^\#\end{array}\right]
&=&  (F+G_uK_{cy}H)\left[\begin{array}{l} 
a\\ a^\#\end{array}\right] \nonumber \\ &&+(G_w+G_uK_{cy}K)
\left[\begin{array}{l}  \mathcal{W} \\ \mathcal{W}^\# 
\end{array}\right] \nonumber \\ &&
 + G_uK_{cw}\left[\begin{array}{l}  \mathcal{\tilde W} \\ \mathcal{\tilde W}^\# 
\end{array}\right]  + G_u
\left[\begin{array}{l}\mathcal{U} \\ \mathcal{U}^\#\end{array}\right]; \nonumber \\
\left[\begin{array}{l} \mathcal{Y} \\\mathcal{Y}^{\#}\end{array}\right]
 &=&
H \left[\begin{array}{l}  a\\
a^\#\end{array}\right]  + K
\left[\begin{array}{l}  \mathcal{W} \\ \mathcal{W}^\# 
\end{array}\right]; 
\end{eqnarray}

\noindent
{\bf Modified Controller:}
\begin{eqnarray}
\label{modifiedcontroller} 
\left[\begin{array}{l} \dot a_c\\ \dot
a_c^\#\end{array}\right]
&=&  F_c\left[\begin{array}{l} 
a_c\\ a_c^\#\end{array}\right] +G_{cw}
\left[\begin{array}{l}   \mathcal{\tilde W}_a \\ \mathcal{\tilde W}_b \\ \mathcal{\tilde W}_a^\# \\
\mathcal{\tilde W}_b^\# 
\end{array}\right]  \nonumber \\ &&+ G_{cy}
\left[\begin{array}{l} 
 \mathcal{Y} \\  \mathcal{Y}^\#
\end{array}\right]; \nonumber \\
\left[\begin{array}{l} 
 \mathcal{U} \\ \mathcal{U}^{\#}
\end{array}\right]
 &=&
H_c
\left[\begin{array}{l}  a_c\\
a_c^\#\end{array}\right]. 
\end{eqnarray}

In order to develop a condition for the physical realizability of the modified controller (\ref{modifiedcontroller}) purely in terms of the dynamics matrices $F_c$, $G_{cy}$, and $H_c$, we 
first apply Theorem \ref{T1} to the augmented controller system (\ref{augcontroller}). From this it follows that (\ref{modifiedcontroller}) is physically realizable if and only if there exists a complex matrix
$\Theta =\Theta^\dagger$  such that 
\begin{eqnarray}
\label{physrealcontroller} && F_c\Theta + \Theta  F_c^\dagger + 
G_cJ G_c^\dagger = 0; ~~
G_c =  -\Theta  \tilde H_c^\dagger J.
\end{eqnarray}
Here, 
\begin{eqnarray*}
G_c &=& \left[\begin{array}{llll}G_{cw1}& G_{cy1}& G_{cw2}& G_{cy2} \end{array}\right];\nonumber \\ 
\tilde H_c &=& \left[\begin{array}{c}H_{c1}\\\tilde H_{c1}\\H_{c2}\\\tilde H_{c2}\end{array}\right];\nonumber \\
G_{cw1} &=& \left[\begin{array}{ll}G_{cw1a} & G_{cw1b}\end{array}\right];\nonumber \\
G_{cw2} &=& \left[\begin{array}{ll}G_{cw2a} & G_{cw2b}\end{array}\right].
\end{eqnarray*}
The conditions (\ref{physrealcontroller}) can be rewritten as
\begin{eqnarray}
\label{physrealcontroller1}
&& F_c\Theta + \Theta  F_c^\dagger + 
G_{cw1a}G_{cw1a}^\dagger+G_{cw1b}G_{cw1b}^\dagger+G_{cy1}G_{cy1}^\dagger\nonumber \\
&&-G_{cw2a}G_{cw2a}^\dagger-G_{cw2b}G_{cw2b}^\dagger - G_{cy2}G_{cy2}^\dagger = 0
\end{eqnarray}
and 
\begin{eqnarray}
\label{physrealcontroller2}
G_{cw1a} &=& -\Theta H_{c1}^\dagger;~~ 
 \left[\begin{array}{ll} G_{cw1b} &  G_{cy1}\end{array}\right]= -\Theta \tilde H_{c1}^\dagger; \nonumber \\ 
G_{cw2a} &=& \Theta H_{c2}^\dagger;~~ 
 \left[\begin{array}{ll} G_{cw2b} &  G_{cy2}\end{array}\right]= \Theta \tilde H_{c2}^\dagger.
\end{eqnarray}
Substituting from (\ref{physrealcontroller2}) into (\ref{physrealcontroller1}), we obtain the following necessary and sufficient condition for the physical realizability of a controller (\ref{modifiedcontroller}) with dynamics matrices $F_c$, $G_{cy}$, and $H_c$:
\begin{eqnarray}
\label{physrealARE}
&&F_c\Theta + \Theta  F_c^\dagger -\Theta \left(H_{c2}H_{c2}^\dagger - H_{c1}H_{c1}^\dagger \right)\Theta \nonumber \\
&&+G_{cy1}G_{cy1}^\dagger - G_{cy2}G_{cy2}^\dagger\nonumber \\
&&+G_{cw1b}G_{cw1b}^\dagger-G_{cw2b}G_{cw2b}^\dagger = 0.
\end{eqnarray}
From this, we can make the following observations. Given a triple $F_c$, $G_{cy}$, $H_c$ and a Hermitian commutation matrix $\Theta$, we can always find matrices $G_{cw1b}$ and $G_{cw2b}$ such that 
\[
G_{cw2b}G_{cw2b}^\dagger-G_{cw1b}G_{cw1b}^\dagger = M
\]
where
\begin{eqnarray*}
M &=& F_c\Theta + \Theta  F_c^\dagger -\Theta \left(H_{c2}H_{c2}^\dagger - H_{c1}H_{c1}^\dagger \right)\Theta \nonumber \\
&&+G_{cy1}G_{cy1}^\dagger - G_{cy2}G_{cy2}^\dagger,
\end{eqnarray*}
and hence, the Riccati equation (\ref{physrealARE}) will be satisfied. Therefore, any controller dynamics defined by a triple $F_c$, $G_{cy}$, $H_c$ can be made physically realizable by the addition of suitable controller noises. This is essentially the result of Lemma 5.6 of \cite{JNP1}. However, the addition of these noises will be detrimental to closed loop performance in both the LQG and $H^\infty$ cases being considered. Also, if the triple $F_c$, $G_{cy}$, $H_c$ is such that the Riccati equation 
\begin{eqnarray*}
&&F_c\Theta + \Theta  F_c^\dagger -\Theta \left(H_{c2}H_{c2}^\dagger - H_{c1}H_{c1}^\dagger \right)\Theta \nonumber \\
&&+G_{cy1}G_{cy1}^\dagger - G_{cy2}G_{cy2}^\dagger =0
\end{eqnarray*}
has a Hermitian solution $\Theta$, then the controller dynamics defined by the triple $F_c$, $G_{cy}$, $H_c$ can be made physically realizable with a minimum number of controller noises and the controller noises $\mathcal{\tilde W}_b$ are not required. This is essentially the main result of \cite{VuP2a}.

We now give a game theory interpretation of the physical realizability condition (\ref{physrealARE}). Indeed, it follows from \cite{DS09} that the Riccati equation (\ref{physrealARE}) will have a Hermitian solution if and only if the following game for the system (\ref{modifiedcontroller}) 
\[
\inf_{\mathcal{Y}(\cdot),\mathcal{\tilde W}_b(\cdot)}\sup_{\mathcal{Y}^\#(\cdot),\mathcal{\tilde W}_b^\#(\cdot)} J
\]
has a finite value for all $T > 0$. Here
\[
J = \int_0^T\left(\begin{array}{l}\|\mathcal{U}\|^2-\|\mathcal{U}^\#\|^2+\|\mathcal{Y}\|^2+\|\mathcal{\tilde W}_b\|^2\\
-\|\mathcal{Y}^\#\|^2 -\|\mathcal{\tilde W}_b^\#\|^2\end{array}\right)dt.
\]
This constraint can be used in place of the constraints (\ref{timedomainPR1}) and (\ref{timedomainPR2}) in the time domain coherent LQG and $H^\infty$ control problems. 
\section{Coherent LQG and $H^\infty$ Control for Annihilation Operator Systems}
In problems of coherent LQG control and coherent $H^\infty$ control for annihilation operator systems, the plant (\ref{plant}) and controller (\ref{controller}) are replaced with the following QSDEs:

\noindent
{\bf Plant:}
\begin{eqnarray}
\label{plantao} 
 d a
&=&  Fadt
+G_{w}d\mathcal{W} + G_{u}d\mathcal{U}; \nonumber \\
d\mathcal{Y}
 &=&
H adt  + K  d\mathcal{W};
\end{eqnarray}

\noindent
{\bf Controller:}
\begin{eqnarray}
\label{controllerao} 
 d a_c
&=&  F_ca_cdt +G_{cw} d\mathcal{\tilde W}+  G_{cy} d\mathcal{Y}; \nonumber \\
 d\mathcal{U}
 &=&
  H_{c}  a_cdt + K_{cw} d\mathcal{\tilde W} + K_{cy} d\mathcal{Y}.
\end{eqnarray}
Also, the augmented controller (\ref{augcontroller}) is replaced with the following QSDEs:
\begin{eqnarray}
\label{augcontrollerao} 
d a_c
 &=&  F_ca_cdt +  \left[\begin{array}{ll}G_{cw}& G_{cy} \end{array}\right]
\left[\begin{array}{l} d\mathcal{\tilde W}\\d\mathcal{Y}\end{array}\right];
\nonumber \\
\left[\begin{array}{l} d\mathcal{U}
\\ d\mathcal{\tilde U} 
\end{array}\right]
 &=&
 \left[\begin{array}{c}H_{c1}\\\tilde H_{c1}\end{array}\right]   a_cdt
+ \left[\begin{array}{llll} K_{cw} & K_{cy}\\
\tilde K_{cw} & \bar K_{cy} 
\end{array}\right] \left[\begin{array}{l} d\mathcal{\tilde W}\\d\mathcal{Y}
\end{array}\right];\nonumber \\
\end{eqnarray}
and the cost output (\ref{costoutput}) is replaced by the equation
\begin{eqnarray}
\label{costoutputao} 
d\mathcal{Z}
 &=&
 C   adt  +  D  d\mathcal{U}.
\end{eqnarray}
The transfer function matrices $\Gamma_P(s)$, $\Gamma_C(s)$, $\tilde \Gamma_C(s)$, and $\Gamma_Z(s)$ are defined as in the previous section for the systems (\ref{plantao}), (\ref{controllerao}), (\ref{augcontrollerao}) and (\ref{costoutputao}). Then, the frequency domain coherent LQG and $H^\infty$ control problems are defined as in the previous section except that Theorem \ref{T4} is used in place of Theorem \ref{T2}. That is, the physical realizability constraint becomes a constraint that the transfer function matrix $\tilde \Gamma_C(s)$ is lossless bounded real. 

In case of annihilation operator systems, the time domain physical realizability conditions (\ref{timedomainPR1}), (\ref{timedomainPR2}) are replaced by the conditions
\begin{eqnarray}
\label{timedomainPR1ao}
\int_0^T\left(\begin{array}{l}
\|\mathcal{U}\|^2 +\|\mathcal{\tilde U}\|^2 
- \|\mathcal{\tilde W}\|^2
-\|\mathcal{Y}\|^2
\end{array}\right)dt
\geq 0
\end{eqnarray}
for all $T > 0$ and all solutions to (\ref{augcontrollerclassical}) with zero initial condition, and 
\begin{eqnarray}
\label{timedomainPR2ao}
\int_0^\infty\left(\begin{array}{l}
\|\mathcal{U}\|^2 +\|\mathcal{\tilde U}\|^2 
- \|\mathcal{\tilde W}\|^2
-\|\mathcal{Y}\|^2\end{array}\right)dt
= 0
\end{eqnarray}
for all solutions to (\ref{augcontrollerclassical}) with zero initial condition satisfying
\[
\int_0^\infty\left(\begin{array}{l}
\|\mathcal{\tilde W}\|^2
+\|\mathcal{Y}\|^2\end{array}\right)dt < \infty.
\]

In the time domain coherent quantum LQG control problem for the case of annihilation operator systems, the cost in (\ref{timedomainlqgcost}) remains the same and physical realizability constraints (\ref{timedomainPR1}), (\ref{timedomainPR2}) are replaced by the constraints (\ref{timedomainPR1ao}), (\ref{timedomainPR2ao}). In the  time domain coherent quantum  $H^\infty$ problem for the case of annihilation operator systems, the game problem in (\ref{Hinfgame}) is replaced by the game problem
\begin{equation}
\label{Hinfgameao}
\inf_{\mathcal{U}(\cdot)} \sup_{\mathcal{W}(\cdot), \mathcal{\tilde W}(\cdot),}
J_\gamma  < \infty 
\end{equation}
subject to the constraints defined by (\ref{timedomainPR1ao}), (\ref{timedomainPR2ao}) where
\[
J_\gamma = \int_0^T\left(\begin{array}{l}\|\mathcal{Z}\|^2 - 
\gamma^2\left(\begin{array}{l}\|\mathcal{W}\|^2
+\| \mathcal{\tilde W}\|^2
\end{array}\right)\end{array}\right)dt.
\]

For the coherent quantum LQG and $H^\infty$ control problems for the case of annihilation operator systems, when we consider the restriction to a strictly proper controller, the modified plant and controller (\ref{modifiedplant}), (\ref{modifiedcontroller}) reduce to the following:

\noindent
{\bf Modified Plant:}
\begin{eqnarray}
\label{modifiedplantao} 
\dot a
&=&  (F+G_uK_{cy}H)a 
+(G_w+G_uK_{cy}K)\mathcal{W}  
\nonumber \\ &&
 + G_uK_{cw}\mathcal{\tilde W}   + G_u
\mathcal{U}; \nonumber \\
\mathcal{Y} 
 &=&
H   a  + K  \mathcal{W}; 
\end{eqnarray}

\noindent
{\bf Modified Controller:}
\begin{eqnarray}
\label{modifiedcontrollerao} 
 \dot a_c
&=&  F_c a_c\ +G_{cw}
\left[\begin{array}{l}   \mathcal{\tilde W}_a \\ \mathcal{\tilde W}_b 
\end{array}\right]  
+ G_{cy} \mathcal{Y} ; \nonumber \\
 \mathcal{U} 
 &=&
H_c a_c,
\end{eqnarray}
where $G_{cw} = \left[\begin{array}{ll}G_{cwa} & G_{cwb}\end{array}\right]$. Also, the Riccati equation condition for physical realizability (\ref{physrealARE}) reduces in this case to the Riccati equation
\begin{eqnarray}
\label{physrealAREao}
F_c\Theta + \Theta  F_c^\dagger +\Theta H_{c}H_{c}^\dagger\Theta +G_{cy}G_{cy}^\dagger+G_{cwb}G_{cwb}^\dagger = 0
\end{eqnarray}
where in this case a solution $\Theta > 0$ is sought. This is a bounded real Riccati equation. From this, it follows that 
a controller with dynamics defined by a triple $F_c$, $G_{cy}$, $H_c$ can be made physically realizable by the addition of suitable controller noises if and only if $F_c$ is Hurwitz and $\|H_c\left(sI-F_c\right)^{-1}\|_\infty \leq 1$. In this case, only the controller noise $\mathcal{\tilde W}_a$ is needed and there is never any advantage in using the controller noise $\mathcal{\tilde W}_b$. This is essentially the result of Theorem 3.2. of \cite{MaP4}. 

\noindent
{\bf Coherent LQG Control for Annihilation Operator Systems:} 
In order to consider the coherent LQG problem for annihilation operator systems, we first assume that all of the noises acting on the plant (\ref{plantao}) and controller (\ref{controllerao}) are purely canonical quantum noises. That is $d\mathcal{W}d\mathcal{W}^\dagger = I dt$ and $d\mathcal{\tilde W}d\mathcal{\tilde W}^\dagger = I dt$; e.g., see \cite{MaP4}. Also, assuming the plant (\ref{plantao}) is physically realizable, it is straightforward to verify using Theorem \ref{T3} that the modified plant corresponding to (\ref{modifiedcontrollerao}) is physically realizable if we ignore the control input $\mathcal{U}$ in the sense that the augmented system 
\begin{eqnarray}
\label{augmodplantao} 
 d a
 &=&  (F+G_uK_{cy}H)adt\nonumber \\
&&+  \left[\begin{array}{ll}G_w+G_uK_{cy} & G_uK_{cw} \end{array}\right]
\left[\begin{array}{l} d\mathcal{W}\\d\mathcal{\tilde W}
 \end{array}\right];
\nonumber \\
\left[\begin{array}{l} d\mathcal{Y}
\\ d\mathcal{\tilde Y}
\end{array}\right]
 &=&
 \left[\begin{array}{l}H\\ \tilde H \end{array}\right] adt  + \left[\begin{array}{ll} K & 0 \\
\tilde K & \bar K 
\end{array}\right] \left[\begin{array}{l} d\mathcal{W}\\d\mathcal{\tilde W}
\end{array}\right]
\end{eqnarray}
is physically realizable. It follows from Theorem \ref{T3} that there exists a matrix $\Theta > 0$ such that 
\begin{eqnarray}
\label{physreal2} && F_a\Theta + \Theta  F_a^\dagger + 
G_a G_a^\dagger = 0; ~~
G_a =  -\Theta  H_a^\dagger,
\end{eqnarray}
where 
\begin{eqnarray*}
F_a &=& F+G_uK_{cy}H; G_a = \left[\begin{array}{ll}G_w+G_uK_{cy} & G_uK_{cw} \end{array}\right]; \nonumber \\
H_a &=& \left[\begin{array}{l}H\\ \tilde H \end{array}\right].
\end{eqnarray*}

Now to consider the coherent LQG problem for the system (\ref{modifiedplantao}), we first ignore the physical realizability constraint on the controller (\ref{modifiedcontrollerao}). In this case, the LQG problem can be solved by using the separation principle; e.g., see \cite{KS72}. In order to do this, we first consider the Kalman Filter for the system (\ref{modifiedplantao}). Noting that $d\mathcal{Y} = L \left[\begin{array}{l} d\mathcal{Y}
\\ d\mathcal{\tilde Y}
\end{array}\right]$ where $L = \left[\begin{array}{ll}I & 0  \end{array}\right]$, we can equivalently consider the Kalman filter for the following system derived from the system (\ref{augmodplantao}):
\begin{eqnarray}
\label{augmodplantao1} 
 d a
 &=&  F_aadt +  G_a \left[\begin{array}{l} d\mathcal{W}\\d\mathcal{\tilde W}
 \end{array}\right];
\nonumber \\
d\mathcal{Y}
 &=&
 LH_a adt  + L \left[\begin{array}{l} d\mathcal{W}\\d\mathcal{\tilde W}
\end{array}\right].
\end{eqnarray}
As in Section 4.3.3 of \cite{KS72}, this Kalman filter is constructed by first finding the  solution $Q \geq 0$ to the Riccati equation
\begin{eqnarray*}
\lefteqn{F_aQ+QF_a^\dagger+G_aG_a^\dagger} \\
&& - (G_a+QH_a^\dagger)L^\dagger(LL^\dagger)^{-1}(G_a+QH_a^\dagger)^\dagger = 0.
\end{eqnarray*}
However, it follows immediately from (\ref{physreal2}) that the matrix $\Theta > 0$ satisfies this equation. Furthermore, the corresponding Kalman gain matrix is given by the formula
\[
K_g = (G_a+QH_a^\dagger)L^\dagger(LL^\dagger)^{-1}
\]
which is equal to zero when $Q=\Theta$ using (\ref{physreal2}). Thus, we can conclude that the Kalman gain is zero and the Kalman state estimate is independent of the output $\mathcal{Y}$. This in turn means that the LQG controller obtained using the separation principle will have a transfer function of zero. However, a zero transfer function automatically satisfies the bounded real condition for physical realizability developed above. Thus, this LQG optimal controller which was derived without regard for the physical realizability constraint, in fact satisfies the physical realizability constraint. That means that this must be the coherent optimal LQG controller. That is, we have proved that the dynamic part of the coherent optimal LQG controller is zero and thus, the total coherent optimal LQG controller must consist purely of a static direct feedthrough term. This leads to the following theorem, which is one of the main results of this paper.

\begin{theorem}
\label{T5}
For any physically realizable annihilation operator plant of the form (\ref{plantao}) and cost (\ref{costoutputao}) with purely quantum noise inputs, then the solution to the corresponding coherent LQG problem will be a purely static controller of the form
\[
d\mathcal{U}
 = K_{cw} d\mathcal{\tilde W} + K_{cy} d\mathcal{Y}.
\] 
\end{theorem}

The above derivation of this theorem also leads to the following corollary which is also one of the main results of this paper. 
\begin{corollary}
\label{C1}
For any physically realizable annihilation operator plant of the form (\ref{plantao}) with purely quantum noise inputs, then the  corresponding Kalman filter dynamics  will be independent of the output $\mathcal{Y}$.
\end{corollary}
Note that  here the  ``Kalman filter'' is defined to be the set of stochastic differential equations constructed via the standard Kalman filter formulas applied to the matrices in the plant model (\ref{plantao}), not the  ``quantum filter'' such as considered in \cite{BHJ07}.

This corollary indicates that for the case under consideration, the output $\mathcal{Y}$ contains no information about the system variables $a$. This conclusion will also hold for any measurements of the quadratures of $\mathcal{Y}$ since such measurements contain less information than $\mathcal{Y}$ itself. 

\noindent
{\bf Coherent $H^\infty$ Control for Annihilation Operator Systems:}
To consider this problem, we assume the plant is physically realizable and that the cost output is of the form 
\begin{equation}
\label{costoutputao1}
d\mathcal{Z} = L \left[\begin{array}{l} d\mathcal{Y}
\\ d\mathcal{\tilde Y} \end{array}\right]
\end{equation}
 where $\mathcal{\tilde Y}$ is the additional output introduced in the augmented plant (see (\ref{augplant})) and $L$ is a matrix whose columns are standard unit vectors. That is, the cost output consists of a collection of physical outputs of the plant. Also, note that $L^\dagger L = I$. 
We first consider the application of the trivial controller $d\mathcal{U} =  d\mathcal{\tilde W}$ to the plant.  It is straightforward to verify that the resulting system will also be physically realizable. Then, it follows that the resulting transfer function $\Gamma(s)$ of the augmented system from $\left[\begin{array}{l} \mathcal{W}\\ \mathcal{\tilde W} \end{array}\right]$ to $\left[\begin{array}{l} \mathcal{Y}\\ \mathcal{\tilde Y} \end{array}\right]$ will be lossless bounded real. Hence, the transfer function $\Gamma_Z(s) = L\Gamma(s) $ from  $\left[\begin{array}{l} \mathcal{W}\\ \mathcal{\tilde W} \end{array}\right]$ to $\mathcal{Z}$ will satisfy:
\[
\Gamma_Z(j\omega)^\dagger \Gamma_Z(j\omega) = \Gamma(j\omega)^\dagger L^\dagger L \Gamma(j\omega) = \Gamma(j\omega)^\dagger  \Gamma(j\omega) = I
\]
for all $\omega \geq 0$. Hence, $\|\Gamma_Z(s)\|_\infty = 1$. 

Now, we consider the application of any physically realizable controller of the form (\ref{controllerao}) to the plant (\ref{plant}). It is straightforward to verify using Theorem \ref{T3} that the resulting closed loop system is physically realizable. It follows that  the resulting transfer function $\tilde \Gamma(s)$ of the augmented system from $\left[\begin{array}{l} \mathcal{W}\\ \mathcal{\tilde W} \end{array}\right]$ to $\left[\begin{array}{l} \mathcal{Y}\\ \mathcal{\tilde Y} \end{array}\right]$ will be lossless bounded real. Hence, the transfer function $\tilde \Gamma_Z(s) = L\tilde \Gamma(s) $ from  $\left[\begin{array}{l} \mathcal{W}\\ \mathcal{\tilde W} \end{array}\right]$ to $\mathcal{Z}$ will satisfy:
\[
\tilde \Gamma_Z(j\omega)^\dagger \tilde \Gamma_Z(j\omega) = \tilde \Gamma(j\omega)^\dagger L^\dagger L \tilde \Gamma(j\omega) = \tilde \Gamma(j\omega)^\dagger  \tilde \Gamma(j\omega) = I
\]
for all $\omega \geq 0$. Hence, $\|\tilde \Gamma_Z(s)\|_\infty = 1$. That is, the trivial controller $d\mathcal{U} =  d\mathcal{\tilde W}$ achieves the same closed loop $H^\infty$ norm as any other physically realizable controller. From this, we obtain the following theorem, which is one of the main results of this paper.

\begin{theorem}
\label{T6}
For any physically realizable annihilation operator plant of the form (\ref{plantao}) and physical cost (\ref{costoutputao1}), then the trivial controller $ d\mathcal{U}
 = d\mathcal{\tilde W}$ will always be an optimal solution to the corresponding coherent $H^\infty$ quantum optimal control problem.
\end{theorem}

\section*{Acknowledgment}
The author acknowledges useful conversations related to the topics of this paper with  Elanor Huntington, Shibdas Roy, and Shanon Vuglar from UNSW Canberra. 


\end{document}